\documentclass[aps,prl,reprint,groupedaddress,amsmath,amssymb]{revtex4-1}

\usepackage{amsthm}
\usepackage{amsmath}
\usepackage{latexsym}
\usepackage{amsfonts}
\usepackage{amssymb}
\usepackage{color}
\usepackage{graphicx}
\usepackage{caption}
\usepackage{subcaption}
\usepackage{tikz}
\usepackage{placeins}
\usepackage[utf8]{inputenc}
\newcommand{\ii}{\mathrm{i}}

\newcommand{\diff}{\mathrm{d}}

\usepackage{braket}

\begin{document}
\title{Revealing the nature of non-equilibrium phase transitions \\with quantum trajectories}

\author{Valentin Link}
\email{valentin.link@tu-dresden.de}

\affiliation{Institut f{\"u}r Theoretische Physik, Technische Universit{\"a}t Dresden, 
D-01062,Dresden, Germany}

\author{Kimmo Luoma}
\email{kimmo.luoma@tu-dresden.de}
\affiliation{Institut f{\"u}r Theoretische Physik, Technische Universit{\"a}t Dresden, 
D-01062,Dresden, Germany}

\author{Walter T. Strunz}
\email{walter.strunz@tu-dresden.de}
\affiliation{Institut f{\"u}r Theoretische Physik, Technische Universit{\"a}t Dresden, 
D-01062,Dresden, Germany}

\date{\today}

\begin{abstract}
{A damped and driven collective spin system is analyzed by using quantum state diffusion. 
This approach allows for a mostly analytical treatment of the investigated non-equilibrium 
quantum many body dynamics, which features a phase transition  
in the thermodynamical limit. 
The exact results obtained in this work, which are free of 
any finite size defects, provide a complete understanding of the model.
Moreover, the trajectory framework gives an intuitive
picture of the two phases occurring,
revealing a spontaneously broken symmetry and allowing for a
qualitative and quantitative characterization of the phases. We determine
exact critical exponents, investigate finite size scaling, and explain a remarkable
non-algebraic behaviour at the transition in terms of torus hopping.} 
\end{abstract}

\pacs{}
\maketitle

\begin{figure}
	\begin{subfigure}{0.44\textwidth} 
		\includegraphics[width=\textwidth]{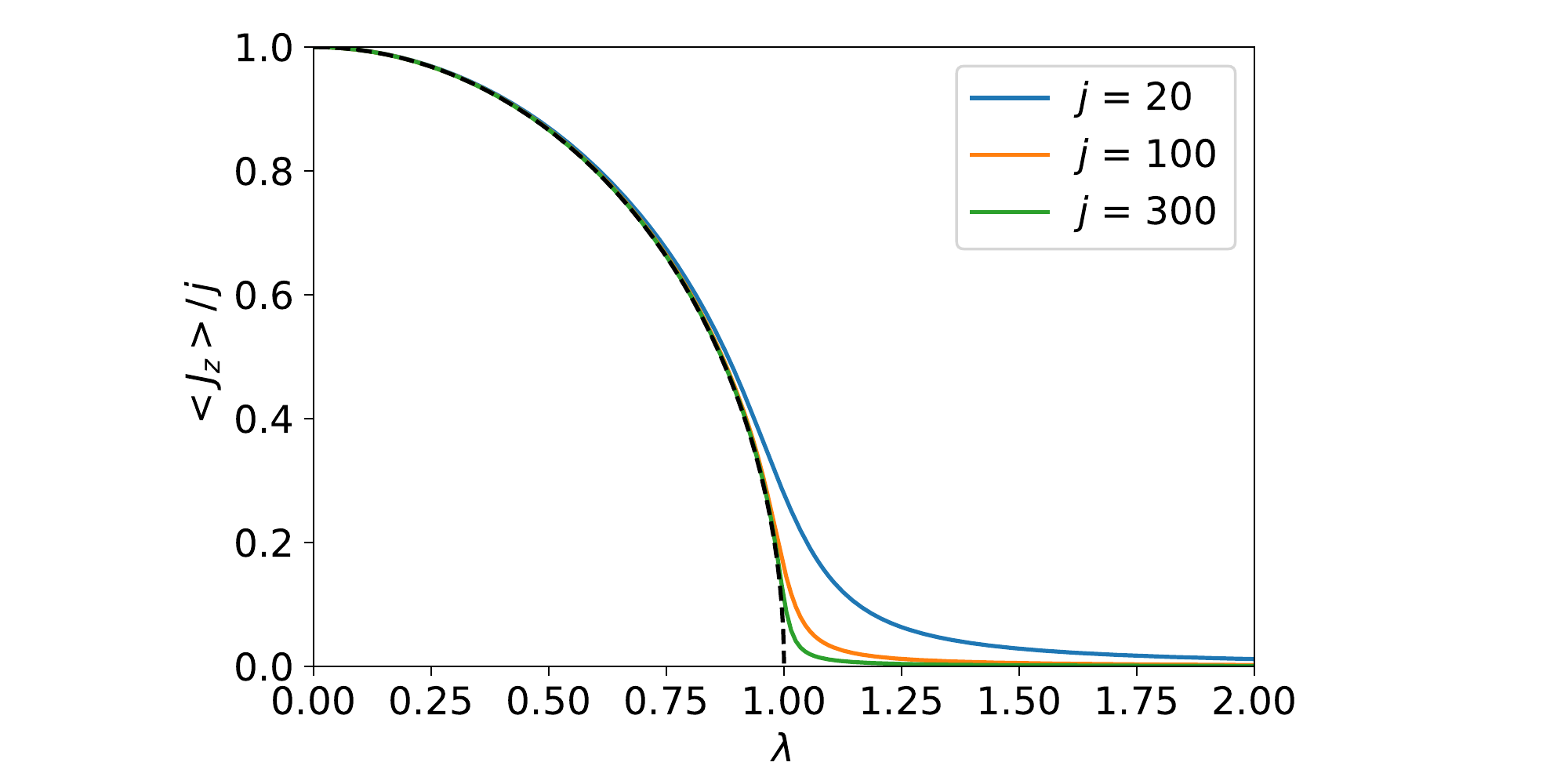}
	\end{subfigure}
   \caption{$J_z$ expectation value in the steady state of master
equation (\ref{eq:lindblad}) as a function of the parameter $\lambda$
for different values of $j$. The dashed line is the asymptotic curve
for $j\rightarrow\infty$ from (\ref{eq:Jzexp}).}
   \label{fig:0}
\end{figure}

\paragraph{Introduction} 
Many important models from the early years of quantum
optics, like the Dicke model \cite{Dicke1954},  have  experienced
renewed interest. The reason being the availability of 
new experimental platforms. First and foremost, the
field of ultra cold atomic gases allows an 
unprecedented level  
of control and
tunability, bringing relatively simple but physically
rich quantum optical models within the reach of current state of the art  
experiments
 \cite{Dimer2007,Torre2013,Gutiregui2018,Hwang2018}. The Dicke model, for 
example, 
can be realized
for a wide range of parameters covering different phases of the system
\cite{Baumann2010,Greiner2002}. Similar models are studied in the 
context of quantum magnetism \cite{Shakirov2016,Ferreira2018,Ribeiro2018}.
Experiments are often performed under
interesting non-equilibrium conditions  where the interplay of driving
and dissipation determines a stationary state of the system in absence
of detailed balance
\cite{Sieberer2016,Sieberer2015,Raftery2014}. The dissipation stems from
interactions with an environment which in many cases cannot be
avoided. If the driven open system exhibits a phase transition upon tuning 
the system
parameters, the non-equilibrium stationary state, rather than the ground state 
of the
Hamiltonian, undergoes a non-analytical change. This poses a formidable 
challenge 
for the theoretical treatment of such a system.
{ The  understanding of phase transitions in driven dissipative quantum many 
body
systems is still developing \cite{Hannukainen2018,Nagy2015,DallaTorre2010}, as 
such  
problems, in general, can be tackled only approximately \cite{Gelhausen2018}. 
}\\ 
{Analytically soluble models containing the relevant physics are highly 
desirable and 
of great value for exploring in great 
detail new phenomena arising in this field. }
A recent trend has been  to
describe driven dissipative systems in the language of non-equilibrium quantum 
field 
theory \cite{Sieberer2016,Torre2013}. 
There, a path integral formalism for Gorini-Kossakowski-Sudarshan-Lindblad 
(GKSL) 
master equations \cite{Gorini1976,*Lindblad1976}, 
conceptually similar to the Keldysh functional integrals, are 
used \cite{Sieberer2016}.  
Such path integral representations of dissipative propagators were earlier 
studied 
in a quantum optics setting in \cite{StrunzJPhysA1997}.\\
Alternatively,  quantum trajectory methods 
from the field of quantum optics \cite{Gisin1992a,Plenio1998}, 
such as quantum state diffusion \cite{Gisin1992a}, also provide an efficient 
and transparent  theoretical framework.  They can contribute to a detailed 
understanding
of non-equilibrium quantum physics, as we will demonstrate in this Letter.  
In particular, the localization property of quantum state diffusion 
in the long time limit \cite{StrunzJPhysA1998} is 
useful for the qualitative and quantitative analysis of non-equilibrium phase 
transitions, since 
this allows a direct  observation of the different
character of distinct phases.
Quantum trajectories  have been used successfully in \cite{Gutiregui2018}
to analyze a generalization of the Dicke model. 
In the present Letter we consider
a model 
similar to the so-called cooperative resonance fluorescence model
introduced in the 1970s \cite{Walls1978,Walls1980,Drummond1978}. The
latter recently received attention because it features a particularly
interesting phase transition which is difficult to characterize
\cite{Iemini2018,Hannukainen2018}.  We here show that our  model can be treated 
exactly
within a quantum trajectory approach. This gives a clear picture of
the different phases, helps to identify a broken
symmetry, allows for an analytical, rather than numerical,  determination of 
critical 
exponents and explains the peculiar critical behaviour.
\paragraph{Model}
We consider the following master equation of GKSL form 
\cite{Gorini1976,*Lindblad1976} for a driven damped spin-$j$ system
\begin{equation}
\begin{split}
 \partial_t \rho=&-\mathrm{i}\omega[J_x,\rho]+\frac{\kappa}{j}\big(J_+\rho 
J_--\frac{1}{2}\{J_-J_+,\rho\}\big)\\
 &+\frac{\kappa}{j}\big(J_z\rho J_z-\frac{1}{2}\{J_z^2,\rho\}\big)\,,
 \end{split}
 \label{eq:lindblad}
\end{equation}
where $J_z$, $J_\pm=J_x\pm \ii J_y$ are {the} spin operators. Apart from the
additional $J_z$ dissipator this coincides with the cooperative
resonance fluorescence model
\cite{Walls1978,Walls1980,Drummond1978,Morrison2008,Schneider2002}. It
can describe $j$ spin-$\frac{1}{2}$ systems undergoing collective
driving and collective damping. 
{Following the proposals in
\cite{Iemini2018,Hannukainen2018} this model could be realized
experimentally with cold atoms. Interestingly, in theoretical models
for engineered atomic spin devices used to describe tunneling
spectroscopy of atomic magnets on metallic surfaces,  
a similar
GKSL dissipator appears naturally \cite{Shakirov2016,Ferreira2018,Ribeiro2018}. 
A phase diagram of this model was examined in
\cite{Ribeiro2018}.}\\
The stationary state of
(\ref{eq:lindblad}) features a phase transition in the thermodynamic
limit $j\rightarrow\infty$ as the parameter
$\lambda=\frac{\omega}{\kappa}$, measuring the relative strength of coherent 
drive 
to dissipation, is changed. For strong damping
$\lambda<1$ the steady state has a finite $J_z$ expectation value. If
instead $\lambda>1$ the $J_z$ expectation value is zero, see Fig.
\ref{fig:0}. \\
It is important to stress that Eq.~(\ref{eq:lindblad}) has 
a discrete symmetry. It is invariant under
mirror reflection given by $J_x\rightarrow -J_x$ followed by complex
conjugation.
The crucial role of this symmetry for the phase transition becomes
apparent when we add a symmetry breaking 
term of the form $\omega_z J_z$ to the Hamiltonian. 
As shown  in \footnote{See Supplemental Material}, the model then features
only a first order transition. 
\paragraph{{Quantum state diffusion approach}} We analyze the model
(\ref{eq:lindblad}) by unraveling the master equation with quantum
state diffusion \cite{Note1}, that is we express the density operator $\rho$ as 
the
average over stochastic pure states \cite{Gisin1992a,Plenio1998}. In
contrast to the cooperative resonance fluorescence model, due to the
additional $J_z$ dissipator, the resulting stochastic Schrödinger
equation preserves spin coherent states for {\it any} $j$, 
that is, for any system size \cite{Note1,Gisin1992}. 
The spin coherent states are defined as
\begin{equation}
 \ket{\mu}=\frac{|\!\ket{\mu}}{\sqrt{\braket{\mu||\mu}}}\,,\qquad 
|\!\ket{\mu}=\mathrm{exp}\big({\mu}J_-\big)\ket{j},
\end{equation}
and the density operator is obtained by averaging over 
{the stochastically evolving} spin coherent states 
$\rho(t)=\mathbb{E}(\ket{\mu(t)}\!\bra{\mu(t)})$ 
where the complex labels $\mu(t)$ are stochastic trajectories  obeying  the 
classical Langevin equation
\begin{equation}
\begin{split}
 \mathrm{d}\mu
 =&\Big(\hspace{-4pt}-\mathrm{i}\frac{\omega}{2}(1- \hspace{-2pt}{\mu^2})
 -\tilde{\kappa}\mu\Big)\mathrm{d}t\hspace{-2pt}
+\hspace{-3pt}\sqrt{\frac{\kappa}{j}}\mu^2\mathrm{d}\xi_+\hspace{-2pt}
-\hspace{-3pt}\sqrt{\frac{\kappa}{j}}\mu\mathrm{d}\xi_z,
 \end{split}\label{eq:ItoLan}
\end{equation}
with complex Ito increments $\mathbb{E}(\mathrm{d}\xi_\alpha)=0$,
$\mathrm{d}\xi_\alpha\mathrm{d}\xi^*_\beta=\delta_{\alpha
\beta}\mathrm{d}t$. A rescaled coupling is introduced as
$\tilde{\kappa}=\kappa(1-\frac{1}{2j})$. That spin
coherent states are preserved is remarkable \cite{Gisin1992}, and reflects the 
localization
property of quantum state diffusion \cite{StrunzJPhysA1998}.  It allows to
solve this model particularly easy. Nevertheless, in the following we
see that this model features qualitatively the same physics as the
cooperative resonance fluorescence model.
Unlike mean-field or semi-classical approaches, Eq.~(\ref{eq:ItoLan})
provides the exact solution of the master equation (\ref{eq:lindblad}) 
for any system size. 
\paragraph{Solution in the thermodynamic limit} 
Neglecting  the noise terms in Eq.~(\ref{eq:ItoLan}) results  in the 
deterministic equation
\begin{equation}
 \dot{\mu}=-\mathrm{i}\frac{\omega}{2}\Big(1-{\mu^2}\Big)-\tilde{\kappa}\mu.
 \label{eq:robust}
\end{equation}
\begin{figure}
	\begin{subfigure}{0.23\textwidth} 
		\includegraphics[width=\textwidth]{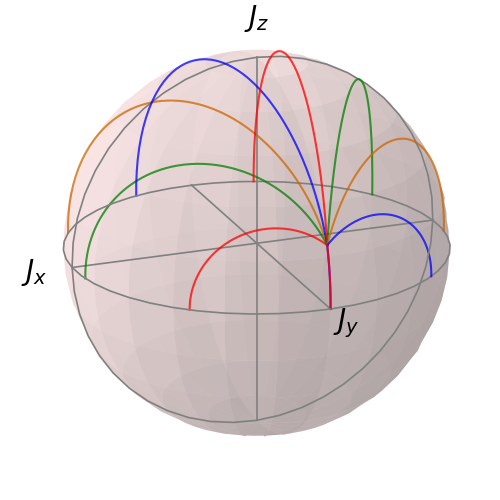}
		\caption{$\lambda=0.95$} 
	\end{subfigure}
	\begin{subfigure}{0.23\textwidth} 
		\includegraphics[width=\textwidth]{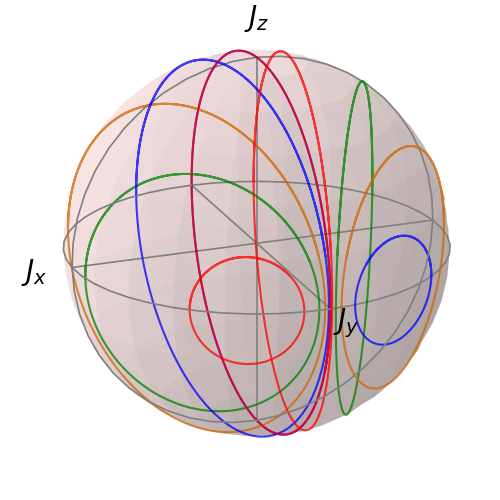}
		\caption{$\lambda=1.05$} 
	\end{subfigure}
   \caption{Deterministic trajectories Eq.~(\ref{eq:robust}) displayed on the
sphere with initial conditions on the equator. In case (a) {$(\lambda<1)$} all
trajectories flow to a stable fixed point. In case (b) {$(\lambda>1)$} 
solutions are
cyclic.}
   \label{fig:2}
\end{figure}
\begin{figure}
	\begin{subfigure}{0.23\textwidth} 
		\includegraphics[width=\textwidth]{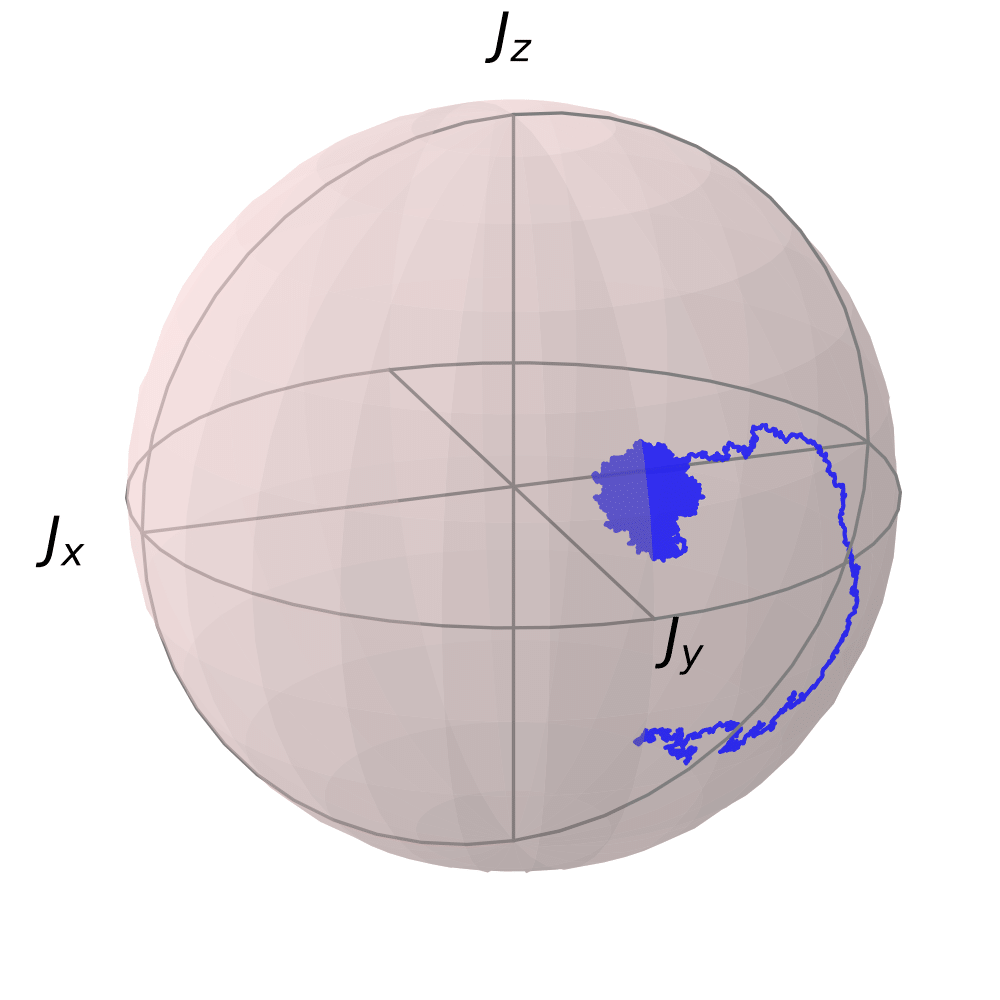}
		\caption{$\lambda=0.95$} 
	\end{subfigure}
	\begin{subfigure}{0.23\textwidth} 
		\includegraphics[width=\textwidth]{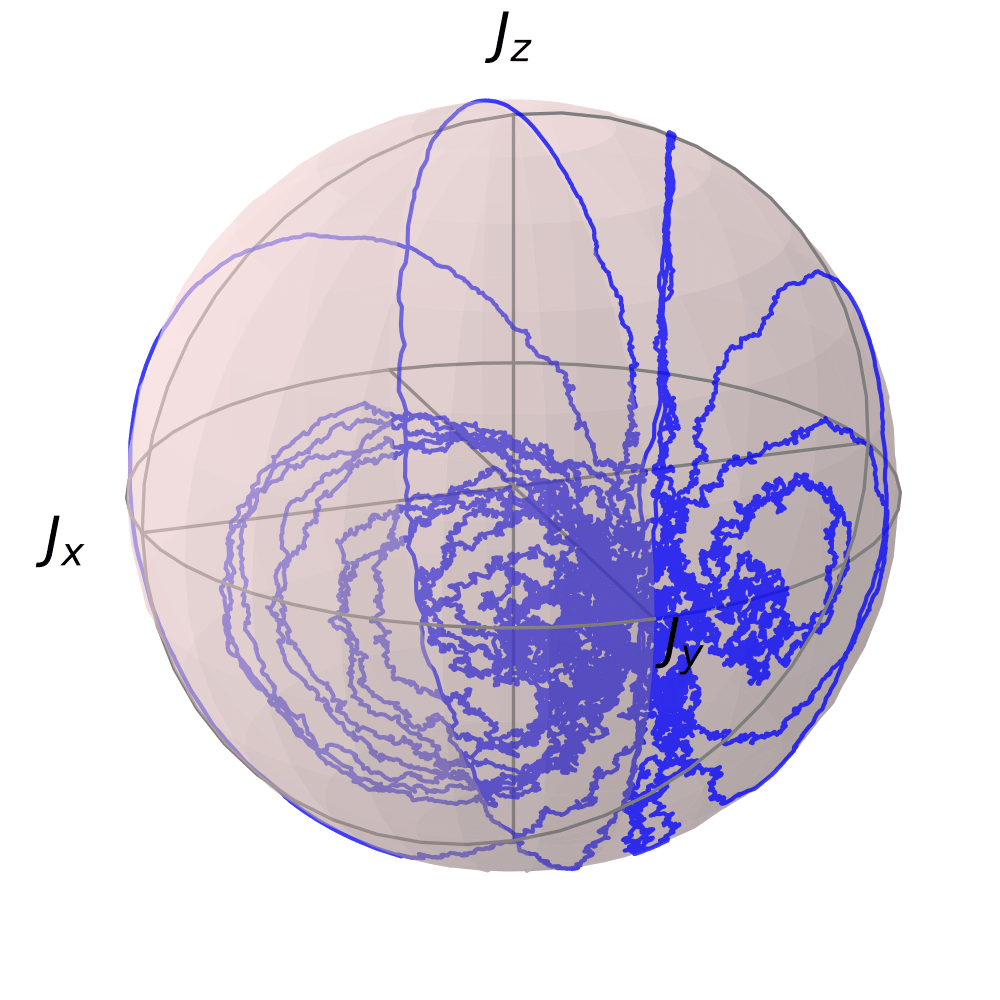}
		\caption{$\lambda=1.05$} 
	\end{subfigure}
   \caption{Example of a noisy trajectory Eq.~(\ref{eq:ItoLan}) for $j=500$. At 
$\lambda<1$
(a) a trajectory starting at the unstable fixed point evolves to the
stable fixed point and remains there. {At $\lambda>1$ (b) a trajectory
starting at the right fixed point hops stocastically between the deterministic 
tori depicted in Fig.~\ref{fig:2}~(b), such that it eventually explores the 
entire phase space.}}
\label{fig:3}
\end{figure}
Since the strength of fluctuations is $\kappa/j$ this is valid  for
times much smaller than $j/\kappa$. In particular, it captures the
thermodynamic limit $j\rightarrow \infty$. The resulting approximate
solutions of Eq.~(\ref{eq:lindblad}) are coherent states with
deterministic label $\mu$ obeying Eq.~(\ref{eq:robust}).  These pure states
are just the so called robust states of
the master equation \cite{Buchleitner2008}. At this point we already
note that for {\it any} finite $j$ the asymptotic state
$\rho(t=\infty)$ of Eq.~(\ref{eq:lindblad}) may nevertheless be highly 
mixed, as will be further
elaborated later. \\
The solution of Eq.~(\ref{eq:robust}) can be
found analytically \cite{Note1}
 \begin{equation}
 \mu_\mathcal{M}(t)=\frac{\mu_-\sqrt{1+\mathcal{M}}\mathrm{ e}^{\ii 
\phi(t)}-\mu_+\sqrt{1-\mathcal{M}}}
{\sqrt{1+\mathcal{M}}\mathrm{ e}^{\ii \phi(t)}-\sqrt{1-\mathcal{M}}},
 \label{eq:mu_analy}
\end{equation}
{with the phase
$\phi(t)=\sqrt{\lambda^2-1}\tilde{\kappa}t+\phi(0)$. $\mathcal{M}$ 
($|\mathcal{M}| \leq 1$) 
and $\phi(0)$ are determined by the initial condition.}
There are two fixed points
\begin{equation}
\mu_\pm=-\mathrm{i}\frac{\tilde{\kappa}}{\omega}\left(1\pm\sqrt{1-\lambda^2}
\right)\,,
\end{equation}
corresponding to $\mathcal{M}=\pm1$.
The most elegant way of displaying the trajectories is to map the
complex label $\mu$ to a point on the sphere via the inverse
stereographic projection given by
\begin{equation}
  \vec{n}=\braket{\mu|\vec{J}|\mu}/j.
\end{equation}
Some of the deterministic trajectories are depicted in Fig.
\ref{fig:2} for two values of $\lambda$.  For $\lambda<1$ all
trajectories flow to the stable fixed point $\mu_-$. 
In contrast, the
solutions are periodic for $\lambda>1$, i.e. {$\mu_{\mathcal{M}}(t)$} traverses 
a
closed torus with period
$T=2\pi(\tilde{\kappa}\sqrt{\lambda^2-1})^{-1}$. As a consequence of the
existence of periodic solutions, the spectrum of the Lindbladian
becomes gapless and the imaginary parts of the eigenvalues are
separated by the fundamental frequency $2\pi /T$, as observed in
\cite{Iemini2018}. {Clearly, the existence of two distinct phases in the
thermodynamic limit is evident.} In the case $\lambda<1$ there is a unique 
stable
(unstable) steady state which is the coherent state
$\rho_-=\ket{\mu_-}\bra{\mu_-}$ ($\rho_+=\ket{\mu_+}\bra{\mu_+}$). For
$\lambda>1$ a steady state can be associated with each torus labeled
with $\mathcal{M}$, by time averaging the cyclic evolution over one
period
\begin{equation}
 \rho_\mathcal{M}=\frac{1}{T}\int_0^{T}\diff t 
\ket{\mu_\mathcal{M}(t)}\bra{\mu_\mathcal{M}(t)}.
 \label{eq:SSm}
\end{equation}
The emergence of periodic solutions has led to the idea that such a
phase transition can be associated with a spontaneous breaking of
continuous time translation symmetry \cite{Iemini2018}. Here, however,
we can clearly see that  the mirror symmetry of the Lindblad generator
(\ref{eq:lindblad}) is spontaneously broken in the $\lambda>1$
phase. The steady states $\rho_\mathcal{M}$ in Eq.~(\ref{eq:SSm}) are not 
mirror symmetric,
since the mirrored state is $\rho_\mathcal{-M}$ \footnote{The mirror
reflected state to a coherent state $\ket{\mu}$ is
$\ket{-\mu^*}$.}. Thus the phase transition happens in presence of an
ordinary symmetry breaking \cite{Ribeiro2018}. Note that all steady
states $\rho_\mathcal{M}$ have a vanishing $J_z$ expectation value and
thus the steady state value of $J_z$ is well defined
\begin{equation}
  \frac{\braket{J_z}}{j}=    \begin{cases}
      \sqrt{1-\lambda^2} &, \lambda<1\\
      0 &, \lambda > 1
    \end{cases}
    \label{eq:Jzexp}
\end{equation}
indicating a second order phase transition. When in the Hamiltonian an
additional symmetry breaking term $\omega_zJ_z$ is present there are
no cyclic solutions and a stable fixed point exists for all
$\lambda$. The system is then no longer { critical at $\lambda=1$
\cite{Note1}.}

\begin{figure}
	\begin{subfigure}{0.44\textwidth} 
		\includegraphics[width=\textwidth]{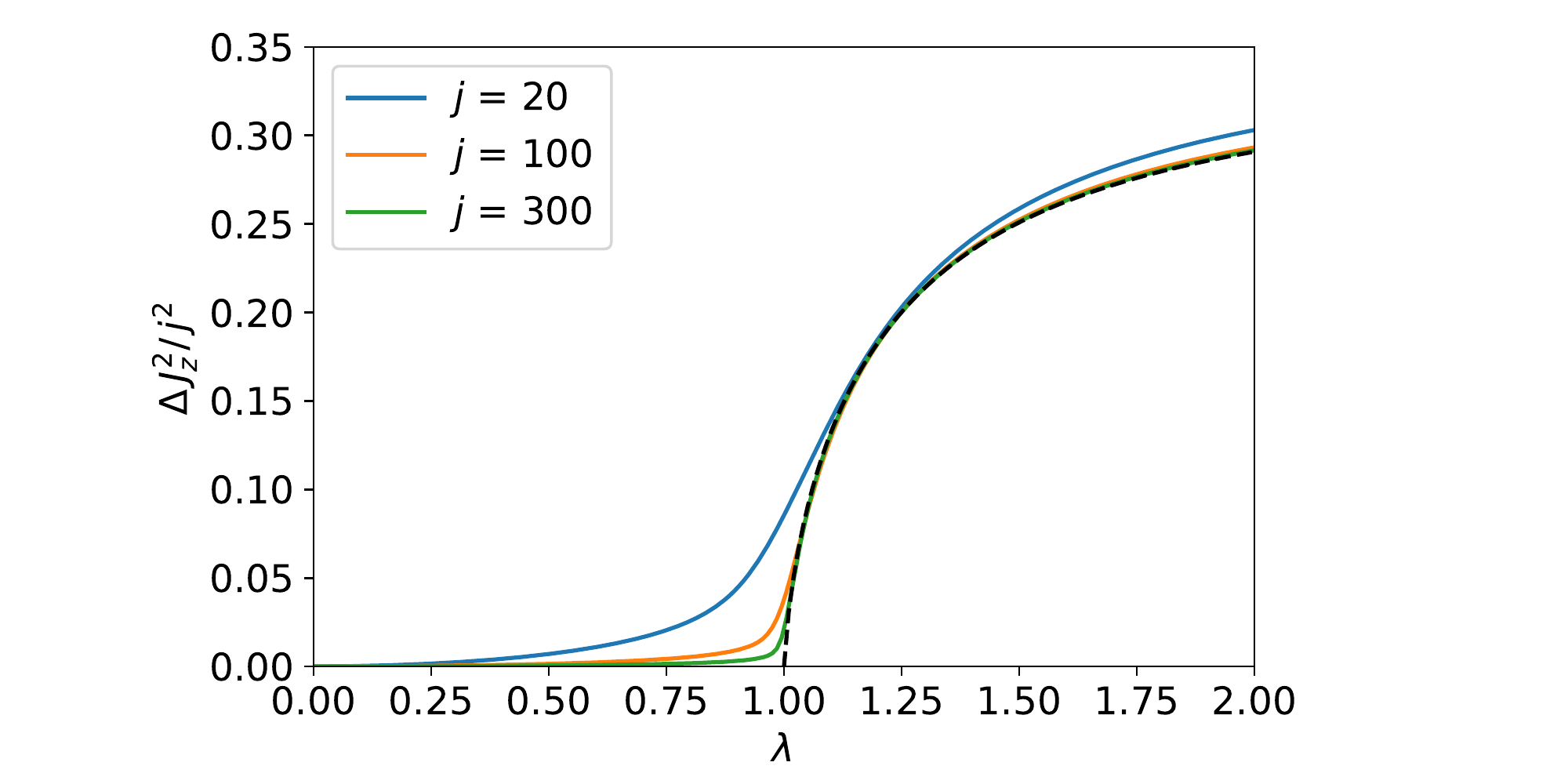}
		\caption{$J_z$ variance.}
	\end{subfigure}
	\begin{subfigure}{0.44\textwidth} 
		\includegraphics[width=\textwidth]{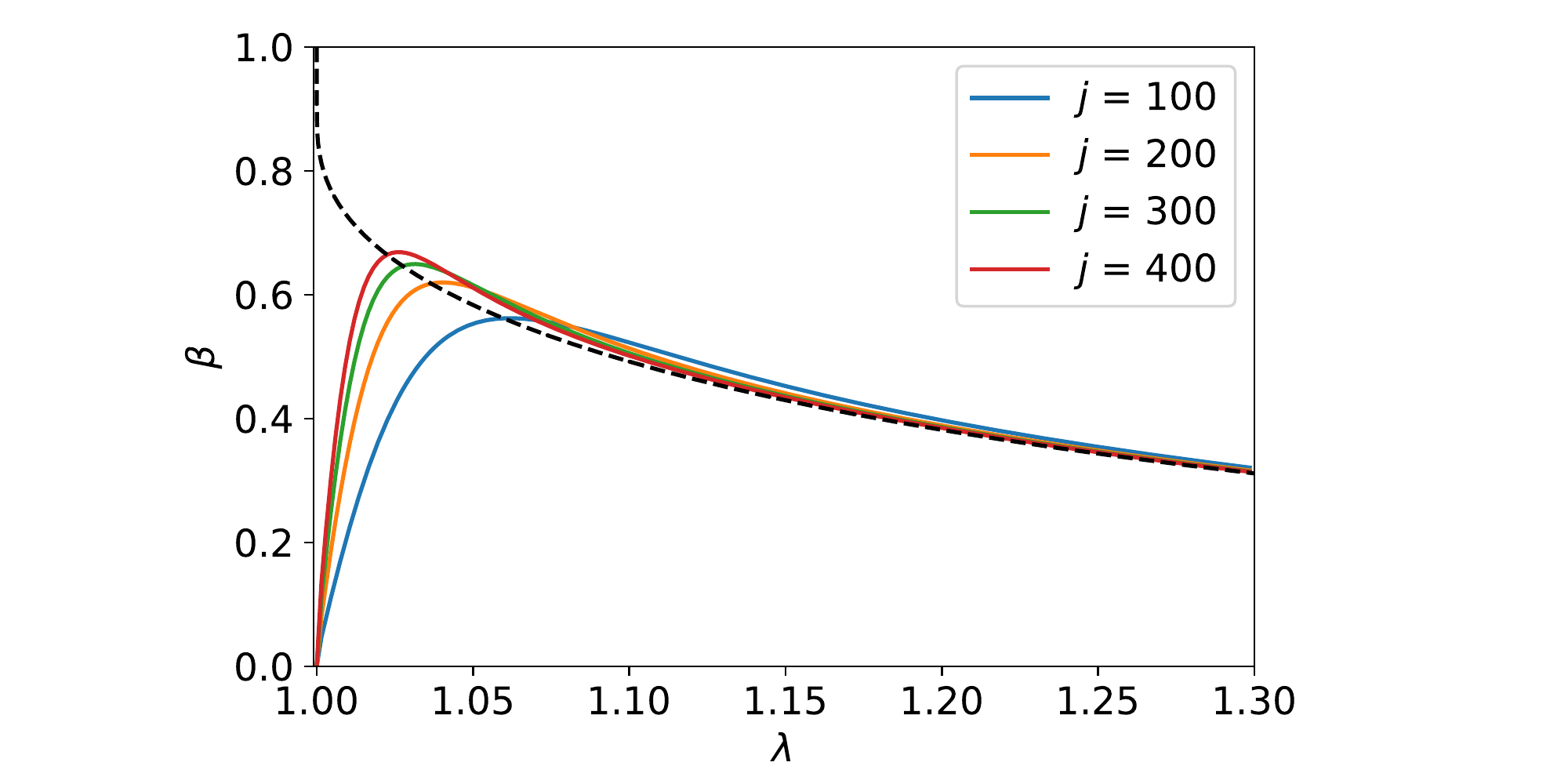}
			\caption{Power law ``exponent'' $\beta$ of the $J_z$ variance.}
	\end{subfigure}

   \caption{Variance of $J_z$ in the steady state of master equation
(\ref{eq:lindblad}) as a function of the parameter $\lambda$ for
different values of $j$ (a). The dashed line is the asymptotic curve for
large $j$ from (\ref{eq:var_est}). {The assignment of a critical exponent 
close to $\lambda=1$ from finite system calculations (b) fails because the 
asymptotic curve does not behave like a power law in the vicinity of the 
transition.}}
\label{fig:1}
\end{figure}

\begin{figure}[htp]
\centering
\includegraphics[width=0.44\textwidth]{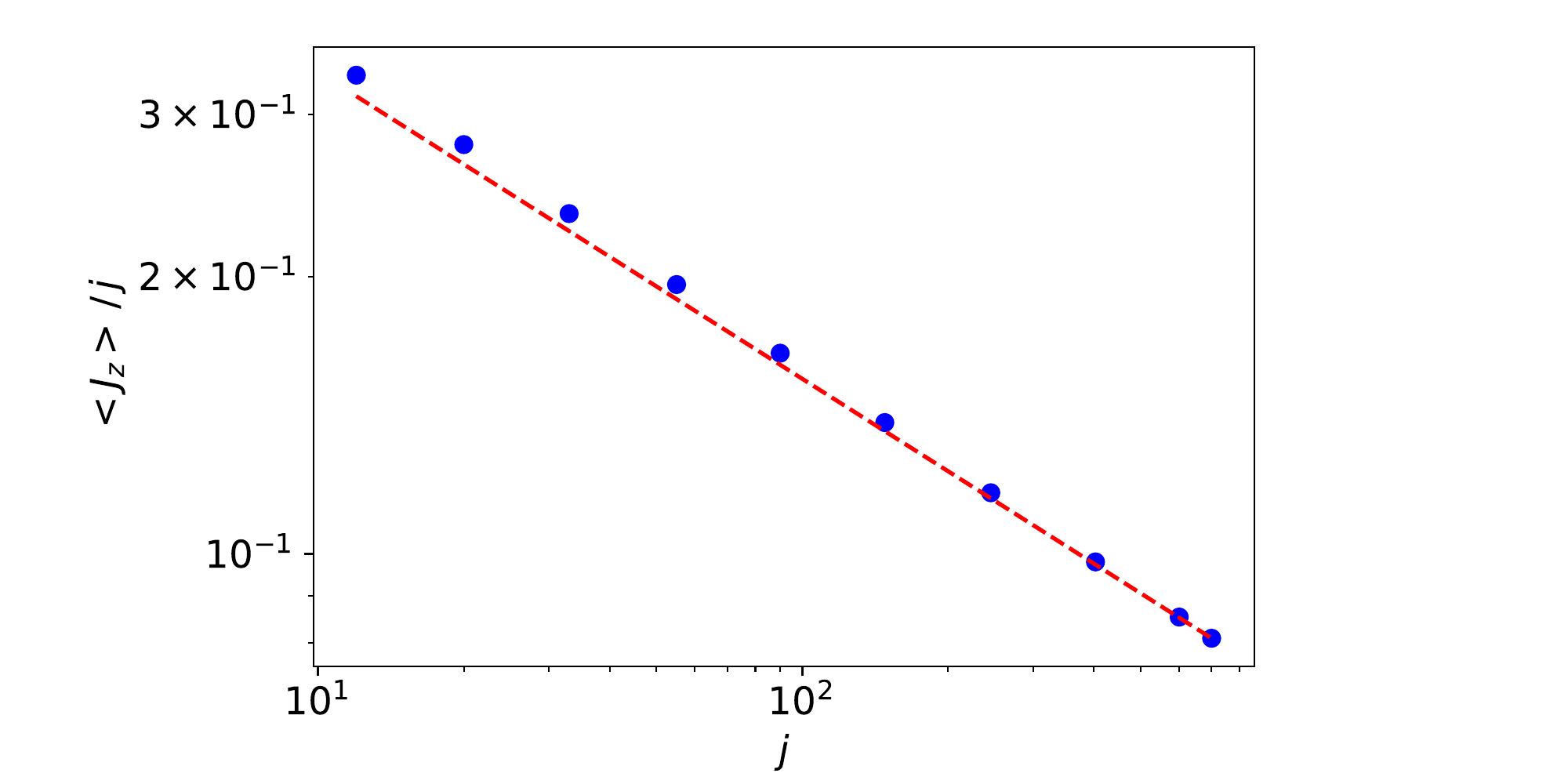}
\caption{$J_z$ steady state expectation value at $\lambda=1$ for different values of $j$. The dashed line has the predicted slope $-1/3$.}
\label{fig:fscaling}
\end{figure}

\paragraph{Finite system size} If the system has a finite size
$j<\infty$ there always exists a unique steady state which is the
asymptotic solution of (\ref{eq:lindblad}). The uniqueness is
seemingly at odds with the previous investigation in the thermodynamic
limit, where for $\lambda>1$ we found a whole family of steady
states. The explanation goes as follows. 
Neglecting the noise terms, as we did in Eq.~(\ref{eq:robust}),
is only valid for times {\it much shorter than} $j/\kappa$. On longer
timescales the noise will lead to a mixing process which {results in} a
unique steady state.  
In Fig.~\ref{fig:3} two quantum
trajectories are displayed in the two regimes. For $\lambda<1$ there
exists a stable fixed point and a weak ($1/j$) noise will only cause small 
fluctuations around this point, see Fig.~\ref{fig:3}~(a). On the
other hand, in the presence of cyclic solutions, when $\lambda>1$, the
noise introduces a hopping between neighboring tori, see Fig.~\ref{fig:3}~(b). 
Torus hopping allows a single trajectory to eventually explore 
the entire phase space.  \\
In Fig.~\ref{fig:1}~(a) the $J_z$ variance of the unique
steady state is displayed for different values of $j$. Clearly, the
variance increases with $\lambda$ and for large system sizes
approaches a curve which is non-analytic at $\lambda=1$. With the
trajectories displayed in Fig.~\ref{fig:3} this can be
understood in an intuitive way.
Moreover, we can even compute analytically the asymptotic curve
displayed in Fig.~\ref{fig:1}~(a). To this
aim it is useful  to switch to the ``action-angle'' variables
$(\mathcal{M},\phi)$ used for the deterministic trajectories
(\ref{eq:mu_analy}).  As elaborated in \cite{Note1},  averaging over the fast 
dynamics of the angle
variable \cite{Carmichael1980,Drummond1978,Kruscha2012} results in a
one dimensional stochastic evolution for the torus label
$\mathcal{M}$, reflecting the slow torus-hopping process. A
stationary distribution for this process can be found analytically
 \begin{equation}
 P(\mathcal{M})=\frac{1}{2}\frac{\sqrt{1+2\lambda^2}}
{\mathrm{tanh}^{-1}\big((1+2\lambda^2)^{-1/2}\big)}\frac{1}{
2\lambda^2+1-\mathcal{M}^2}.
\end{equation}
{The unique steady state resulting from the mixing induced by 
the torus hopping  is}
\begin{equation}
 \rho_{SS}=\int_{-1}^1\diff\mathcal{M}P(\mathcal{M})\rho_\mathcal{M},
\end{equation}
with $\rho_\mathcal{M}$ from (\ref{eq:SSm}).  Naturally,  this unique state 
does have the same
symmetries as the Lindblad generator, since
$P(\mathcal{M})$ is a symmetric distribution. 
We note in passing that the 
additional $J_z$ dissipator  of the present model leads to a different 
distribution of tori compared to the one
derived by Carmichael using a Glauber $P$-distribution technique
in \cite{Carmichael1980}. The $J_z$ variance of $\rho_{SS}$ can be given in 
closed form,
\begin{equation}
\begin{split}
 \frac{\Delta 
J_z^2}{j^2}=(\lambda^2-1)\left(\frac{\mathrm{tanh}^{-1}\big(\sqrt{3}
(1+2\lambda^2)^{ -1/2}\big)}{\sqrt{3}\,\mathrm{ 
tanh}^{-1}\big((1+2\lambda^2)^{-1/2}\big)}-1\right),
 \label{eq:var_est}
 \end{split}
\end{equation}
displayed as the dashed { line} in Fig.~\ref{fig:1}~(a).
{Interestingly, this function does not behave like a power law for
$\lambda$ close to one. With $\lambda = 1+\varepsilon$ we find that
the asymptotic curve
behaves as $-\varepsilon\ln\varepsilon$ \cite{Note1}. This behaviour 
is hard to find from extrapolating finite size calculations, as can be seen in 
Fig.~\ref{fig:1}~(b). 
There we plot ``exponents'' $\beta$ that one would associate to the limiting 
curve by assuming 
a power law, estimated for different values of {$\lambda=1+\varepsilon$}.
Clearly we see from the exact curve that
close to $\lambda=1$ no exponent can be
associated, and the peculiar behaviour $\beta \simeq 
1+\frac{1}{\ln(\varepsilon)}$ 
with an infinite negative slope at the critical point follows \cite{Note1}.

In the thermodynamic limit different steady
states (\ref{eq:SSm}) do not have the same $J_z$ variance, and thus this 
quantity is
not well defined. As a consequence, this observable does not have a
universal power law scaling close to the transition point -- see \cite{Note1} 
for more details. 
Fig.~\ref{fig:1}~(b) manifests also the non-commutativity of the two types
of limiting procedures involved. The non universal behaviour is a finite system 
size effect 
which  emerges when first
the stationary state is found and only afterwards the limit of large system 
size is taken.
In order to correctly understand the phase transition, the order of limits has 
to 
be interchanged. Neglecting the $1/j$-terms, as in (\ref{eq:robust}), 
corresponds to first
taking the strict thermodynamic limit leading to the family of steady states 
(\ref{eq:SSm}). 
Naively computing the steady state of (\ref{eq:lindblad}) for finite $j$ 
results  always 
in a unique steady state and the broken symmetry is not revealed. }
\paragraph{Critical exponents and finite size scaling}
Clearly, Eq.~(\ref{eq:Jzexp}) gives the critical exponent
$1/2$ for the $J_z$ expectation value. From the exact solution in the
thermodynamic limit (\ref{eq:mu_analy}) we can just as easily read off
the dynamical critical exponent. Since there is only a single time
scale $\xi=|1-\lambda^2|^{-1/2}\kappa^{-1}$ the dynamical exponent is
also $1/2$. With the exact Ito-equations at hand it is also possible
to find finite size scaling exponents at $\lambda=1$ by using a 
renormalization scheme as in \cite{Torre2013,Hwang2018}.  
{Numerically we observe a finite size power law scaling at the transition point, 
see
Fig.~\ref{fig:fscaling}. To obtain the exponent analytically we}  set
$\nu=\mu+\ii$ and the evolution equation (\ref{eq:ItoLan}) is rewritten as 
\begin{equation}
 \diff \nu=\ii\frac{\tilde{\kappa}}{2}\nu^2\diff 
t+\sqrt{\frac{\kappa}{j}}(\nu-\ii)\diff\xi_z+\sqrt{\frac{\kappa}{j}}
(\nu-\ii)^2\diff \xi_+\,.
\end{equation}
Under a rescaling of time, close to the transition, $\nu$ has power law scaling. 
 
Indeed, the noiseless equation is invariant under
the transformation
\begin{equation}
 \begin{split}
  t\rightarrow at \,,\qquad
  \nu\rightarrow \frac{\nu}{a}\,.
 \end{split}
\end{equation}
This transformation changes the evolution equation to \footnote{Ito noise 
increments scale as $\diff\xi\rightarrow \sqrt{a}\diff\xi $}
\begin{equation}
  \diff \nu=\ii\frac{\tilde{\kappa}}{2}\nu^2\diff 
t+a^{3/2}\sqrt{\frac{\kappa}{j}}(\frac{\nu}{a}-\ii)\diff\xi_z+a^{3/2}\sqrt{\frac
{\kappa}{j}}
(\frac{\nu}{a}-\ii)^2\diff \xi_+\,.
\nonumber
\end{equation}
For the theory (with fluctuations) to be scale invariant at low frequencies, $j$ 
must scale as \cite{Torre2013,Hwang2018}
\begin{equation}
 \frac{\kappa}{j}\rightarrow \frac{1}{a^3}\frac{\kappa}{j}\,.
\end{equation}
Then all noise terms containing $\nu/a$ become irrelevant under renormalization 
and the resulting low frequency theory with constant noise is scale invariant. 
The finite-$j$ scaling of $\nu$ is found to be
$(\kappa/{j})^{-\frac{1}{3}}\nu\sim 1$.
With $\mu=\nu-\ii$,  we conclude that for large $j$ 
\begin{equation}
 \braket{\mu|J_z|\mu}/j\sim\nu+\mathcal{O}(\nu^2)\approx \nu\sim 
\Big(\frac{\kappa}{j}\Big)^{\frac{1}{3}}\,.
\end{equation}
As seen in Fig.~\ref{fig:fscaling}, numerically accessible values match this 
law quite well.

\paragraph{Conclusions}
Quantum state diffusion
for non-equilibrium quantum dynamics allows us to obtain 
exact analytical results for a 
driven dissipative many body quantum system featuring a phase transition. 
We are able to identify the symmetries of the different phases emerging in the thermodynamical
limit, and in this way we can reveal the symmetry breaking associated with the phase transition.
For these findings the trajectory picture proves insightful. Moreover, it
provides an
elegant shortcut to an exact analytical treatment.
The physics in the trajectory framework can be  
characterized by two time scales, $T$ and $j/\kappa$, 
which describe the deterministic evolution of the open system and the { fluctuations},
respectively. In the thermodynamic limit $(j\to\infty)$ the relaxation time
scale is infinitely large and the system does not relax, resulting in  the degeneracy
of the stationary state. Even though we here consider a solvable model in terms 
of spin coherent states, quantum state diffusion
trajectories would reveal the underlying character of the phases also  
for finite $j$ and 
for non-linear Hamiltonians. These results strongly support the idea that 
quantum trajectories help to unravel non-equilibrium phenomena in 
many body quantum dynamics.

\paragraph{Acknowledgments}
It is a pleasure to thank Holger Kantz and Konrad Merkel for discussions and 
advice. 
V.L. acknowledges support from the
International Max Planck Research School (IMPRS) of MPIPKS Dresden.

\bibliography{bib.bib}
\end{document}